\documentstyle[aps,prl,twocolumn,floats]{revtex}

\preprint{DAMTP-2000-21}
\date{June 26, 2000}
\tighten

\begin{document}
\draft
\newcommand\lsim{\mathrel{\rlap{\lower4pt\hbox{\hskip1pt$\sim$}}
    \raise1pt\hbox{$<$}}}
\newcommand\gsim{\mathrel{\rlap{\lower4pt\hbox{\hskip1pt$\sim$}}
    \raise1pt\hbox{$>$}}}


\title{Primordial Adiabatic Fluctuations from Cosmic Defects}

\author{P. P. Avelino${}^{1,2}$\thanks{
Electronic address: pedro\,@\,astro.up.pt} and
C. J. A. P. Martins${}^{3}$\thanks{Also at C.A.U.P.,
Rua das Estrelas s/n, 4150 Porto, Portugal.
Electronic address: C.J.A.P.Martins\,@\,damtp.cam.ac.uk}
}

\address{${}^1$ Centro de Astrof\'{\i}sica, Universidade do Porto\\
Rua das Estrelas s/n, 4150-762 Porto, Portugal}

\address{${}^2$ Dep. de F{\' \i}sica da Faculdade de Ci\^encias da 
Univ. do Porto\\
Rua do Campo Alegre 687, 4169-007 Porto, Portugal}

\address{${}^3$ Department of Applied Mathematics and Theoretical Physics\\
Centre for Mathematical Sciences, University of Cambridge\\
Wilberforce Road, Cambridge CB3 0WA, U.K.}

\maketitle
\begin{abstract}

{In the context of ``two-metric'' theories of gravity
there is the possibility that cosmic defects will produce a spectrum
of primordial adiabatic density perturbations. This will happen when the
speed characterising the defect-producing scalar field is much larger
than the speed characterising gravity and all standard model
particles. This model will exactly
mimic the standard predictions of inflationary models, with the exception
of a small non-Gaussian signal which could be detected by future experiments.
We briefly discuss defect evolution in these scenarios and analyze their
cosmological consequences.}

\end{abstract} 
\pacs{PACS number(s): 98.80.Cq, 04.50.+h, 98.65.Dx, 98.70.Vc}

\section{Introduction}
\label{secintro} 

Cosmology is entering a crucial stage, where a growing
body of high-precision data will allow us to determine a number of
cosmological parameters, and to identify the mechanism
that produced the ``seeds'' for the structures we observe today \cite{kolb}.
There are currently two classes of models that could be responsible
for these---topological defect \cite{vsh} and
inflationary \cite{linde} models.
The main difference between them is related to causality.
Initial conditions for the defect network
are set up on a Cauchy surface that is part of the standard history of the universe. Hence, there will not be
any correlations between quantities defined at any two spacetime points
whose backward light cones do not intersect on that surface.
Inflation pushes this surface to much earlier times,
and if the inflationary epoch is long enough there will be
essentially no causality constraints.
This can also be seen by noting that inflation can be
defined as an epoch when the comoving Hubble
length decreases. It starts out very large, and perturbations
can be generated causally. Then inflation forces this length to
decrease enough so that, even though it grows again after inflation ends,
it's never as large (by today) as the pre-inflationary era value. Once
primordial fluctuations are produced they can simply freeze in comoving
coordinates and let the Hubble length shrink and then (for small enough
scales) grow past them.

As a step towards identifying the specific model that operated in
the early universe, one would like to determine which of the two
mechanisms above was involved. The presence of
super-horizon perturbations might seem a good enough test, but
this is not the case: in defect models (as well as open or $\Lambda$-models)
significant contributions are generated after the epoch of
last scattering due to the integrated Sachs-Wolfe effect.
The presence of the `Doppler peaks' on small angular scales \cite{gun}
is also not ideal: Turok \cite{turok}
has shown that a causal scaling source can be constructed so as
to mimic inflation and reproduce its contribution
to the CMB anisotropies.
This source is constructed ``by hand'', and there is
no attempt to provide a framework in which
it could be realized. In any case, it
shows that inflationary predictions are not as unique as one might
think. We should also mention, however, that a nice argument due to
Liddle \cite{hor} (see also \cite{also}) shows that the existence of
adiabatic perturbations on
scales much larger than the Hubble radius implies that either inflation 
occurred in the past, the perturbations were there as initial 
conditions, or causality (or Lorentz invariance) is violated.
On the other hand, it is also possible to construct
``designer inflation'' models \cite{desi} that would have no secondary
Doppler peaks, although these suffer from analogous caveats and they would
still be identifiable by other means \cite{liddle,alex}.

Finally, there are Gaussianity tests. There have been recent claims of
a non-Gaussian component in the CMB \cite{gauss1} (but see also \cite{gauss2}). Defects will generally produce non-Gaussian fluctuations
on small enough scales \cite{ASWA2}, whereas the simplest inflationary models
produce Gaussian ones. It's possible to build inflationary models
that produce, {\em eg.} non-Gaussianity with a chi-squared
distribution \cite{isoc}, but if one
found non-Gaussianity in the form of line discontinuities, then it is hard
to see how cosmic strings could fail to be involved.

This discussion shows that although defect and inflationary models
have of course a number of distinguishing characteristics, there is a
greater overlap between them than most people would care to admit.
It is also easy to obtain models where both defects and
inflation generate density fluctuations \cite{acm2,cb}.
The aim of this letter is to present a further example of this overlap.
We discuss a model where the primordial fluctuations are generated by a defect
network, but are nevertheless very similar to a standard inflationary model.
The only difference between these models and the standard inflationary
scenario will be a small non-Gaussian component.
A detailed discussion will be presented in a forthcoming
publication \cite{fut}.

\section{The model}
\label{model} 

Our model follows the recent work on so-called `varying speed of light'
theories \cite{mof1,abm,AM2,dru,AMR,bass,mof2},
and more particularly the spirit of `two-metric'
theories \cite{mof1,dru,bass,mof2},
having two natural speed parameters, say $c_\phi$ 
and $c$; the first is relevant for the dynamics of the scalar field which
will produce topological defects, while 
the second is the ordinary speed of light that is relevant for gravity and
all standard model interactions.

We assume that $c_\phi \gg c$ so
that the correlation length of the network of topological defects will be
much greater than the horizon size.
We could, in analogy with \cite{mof2,bass}, define our
effective theory by means of an action, and postulate a relation
between the two metrics. However, this is not needed for the basic point we're
discussing, so we leave it for a future publication \cite{fut}.
We concentrate on the case of
cosmic strings, whose dynamics and evolution are better known than
those of other defects \cite{vsh,ms1,ms2,thesis}
although much of what we will discuss will apply to them as well.
Note that $c_\phi$ could either be a constant (say
$[g_\phi]_{00}=(c_\phi^2/c^2) g_{00}$)
or, as in \cite{bass} one could set up a model such that
the two speeds are equal at very early and at recent times,
and between these two epochs there
is a period, limited by two phase transitions, where $c_\phi \gg c$. As will
become clear below, the basic mechanism will work in both cases, although
the observational constraints on it will be different for each
specific realization.

The string network evolution is qualitatively
analogous to the standard
case \cite{vsh,ms1,ms2,thesis}, and in particular
a ``scaling'' solution will be reached
after a relatively short transient period.
The long-string characteristic length (or ``correlation length'')
$L$ will evolve as $L=\gamma c_\phi t$, 
with $\gamma={\cal O}(1)$, while the string RMS velocity will
obey $v_\phi=\beta c_\phi$, with $\beta<1$.
Note, however, that there are some differences
relative to the standard scenario.
The first one is obvious: if $c_\phi \gg c$, the string
network will be outside the horizon, measured in the usual way. Hence
these defects will induce fluctuations when they are
well outside the horizon, thus avoiding causality constraints
Note that compensation now acts outside the `$c_\phi$-horizon'.
We expect the effect of gravitational back-reaction
to be much stronger than in the standard case \cite{model,bass}.
The general effect of the back-reaction is to reduce the
scaling density and velocity of the network relative to the standard
value \cite{model}. Thus we should expect fewer 
defects per ``$c_\phi$-horizon'', than in the standard case.
However, despite this strong
back-reaction, strings will still move relativistically. It can be
shown \cite{model} that although back-reaction can slow strings down by
a measurable amount, only friction forces \cite{ms2,thesis} can
force the network into a strong non-relativistic regime.
Thus we expect $v_\phi$ to be somewhat lower than $c_\phi$, but still larger
than $c$.
Only in the case of monopoles, which are point-like,
one would expect the defect velocities
to drop below $c$ due to graviton radiation \cite{bass}.
This does not happen for extended objects, since their tension
naturally tends to make the dynamics take place with a characteristic
speed $c_\phi$ \cite{carter}.
This is actually crucial: if the network was completely
frozen while it was outside the horizon (as in standard scenarios \cite{acm2})
then no significant perturbations would be generated.

A third important aspect
is that the the symmetry breaking scale, say $\Sigma$,
which produces the defects can be significantly lower than the GUT scale,
as density perturbations can grow for a longer time than usual.
The earlier the defects are formed, the lighter they could be. Proper
normalization of the model will produce a further constraint on $\Sigma$.
Finally, in the case
where $c_\phi$ is a time-varying quantity which only departs from $c$
for a limited period,
the defects will become frozen and start to fall inside the horizon
after the second phase transition. Here we require that
the defects are sufficiently outside the horizon and are relativistic
when density fluctuations in the observable scales are generated.
This will introduce additional constraints on model parameters,
notably on the epochs at which the phase transitions take place.

\section{Cosmological consequences}
\label{growth} 

In the synchronous gauge, the linear evolution equations for radiation and 
cold dark matter perturbations, $\delta_r$ and $\delta_m$, in a 
flat universe with zero cosmological constant are
\begin{eqnarray}
  \ddot \delta_m + {\dot a \over a} \dot \delta_m -
  {3 \over 2}\Big({\dot a \over a}\Big)^2 \, \left({a
      \delta_m + 2 a_{eq} \delta_r
      \over a + a_{eq}}\right)  = 4 \pi G 
  \Theta_+,
  \label{one}\\
  \ddot \delta_r - {1 \over 3} \nabla^2 \delta_r
  - {4 \over 3}
  \ddot \delta_m = 0\, ,
  \label{two}
\end{eqnarray}
where $\Theta_{\alpha \beta}$ is the energy-momentum tensor of the external source, $\Theta_+ =  \Theta_{00} +\Theta_{ii}$, $a$ is the scale factor,
``{\it eq}'' denotes the epoch of radiation-matter equality,
and a dot represents a derivative with
respect to conformal time. We will consider the growth of super-horizon 
perturbations with $c k \eta \ll 1$. Then eqn. (\ref{one}) becomes:
\begin{equation}
 \ddot \delta_m + {\dot a \over a} \dot \delta_m -
  {1 \over 2}\Big({\dot a \over a}\Big)^2 \, \left({3a
      + 8 a_{eq} 
      \over {a + a_{eq}}}\right) \delta_m  = 4 \pi G 
  \Theta_+\, ,
  \label{three}
\end{equation}
and $\delta_r=4\delta_m/3$.
Its solution, with initial conditions 
$\delta_m =0$, ${\dot \delta_m}=0$ can be written as
\begin{equation}
\delta^S_{m}({\bf x},\eta) = 4 \pi G \int_{\eta_i}^\eta d\eta' \,
\int d^3x' {\cal G}(X;\eta,\eta') \Theta_{+}({\bf x'},\eta')\, ,
\end{equation}
\begin{equation}
{\cal G}(X;\eta,\eta') = {1 \over 2 \pi^2} \int_0^\infty \, \widetilde {\cal
G}(k;\eta,\eta') {\sin k X \over k X} k^2 dk\, . 
\end{equation} 
Here $X=|{\bf x} -{\bf x'}|$ and `S'
indicates that these are the `subsequent' fluctuations, according to the
notation of \cite{VS}, to be distinguished from `initial' ones.

We are interested in computing the inhomogeneities at late times in
the matter era. When
$\eta_0 \gg \eta_{eq}$, the Green functions are dominated by the growing mode,
$\propto a_0/a_{eq}$, so the function we would like to solve for is \cite{VS}
\begin{equation}
T(k;\eta) = \lim_{\eta_0/\eta_{eq} \to \infty} {a_{eq} \over a_0}
\widetilde {\cal G}(k,\eta_0,\eta)\, .
\label{transfer}
\end{equation}
Consider the growth of  super-horizon perturbations, for which the transfer function can be written \cite{VS}
\begin{equation}
T(0;\eta) = \frac{\eta_{eq}}{10(3-2{\sqrt 2}) \eta}\, .
\label{T0}
\end{equation}
Linear perturbations induced by defects are
the sum of initial and subsequent perturbations:
\begin{eqnarray}
\delta_m(k;\eta_0) &=& \delta_m^I(k;\eta_0) + \delta_m^S(k;\eta_0)
\cr\cr &=& 4 \pi G (1 + z_{eq})
\int_{\eta_i}^{\eta_0} \, d\eta\, T_c(k;\eta) 
\widetilde\Theta_{+}(k;\eta)\, ,
\end{eqnarray}
where $\eta_i$ is the epoch of defect formation.
The transfer function for the subsequent perturbations, those
generated actively, was obtained in eqn. (\ref{T0}) for super-horizon 
perturbations with $c k \eta_0 \ll 1$. To include 
compensation for the initial
perturbations, $\delta_m^I$, we make the
substitution $ T_c(k;\eta) = \Big(1 + (k_c/k)^2 \Big)^{-1} \, T(k;\eta)$,
where $k_c \propto (c_\phi \eta)^{-1}$ is a long-wavelength cut-off at the compensation scale. 
This results from the fact that defect perturbations cannot propagate with a velocity greater than $c_\phi$. 
For $(c_\phi \eta_0)^{-1}  \ll k \ll (c_\phi \eta_i)^{-1}$ the analytic expression for the power spectrum of
density perturbations induced by defects is 
\begin{equation}
P(k)=16 \pi^2 G^2 (1+z_{\rm eq})^2 \int_0^{\infty}d\eta {\cal F}(k,\eta)|T_c(k,\eta)|^{2}\, ,
\label{pspec}
\end{equation}
where $ {\cal F}(k,\eta)$ is the structure function which can be obtained 
directly from the unequal time correlators \cite{VS,WASA,AS}. 
It can be shown\cite{WASA} that for a scaling 
network ${\cal F}(k,\eta)= {\cal F}(k \eta)$ which, combined with the above
relations gives 
\begin{equation}
P(k) \propto \int_0^{\infty}d\eta {\cal S}(k \eta) / \eta^2 \propto k\, 
\label{pspec2}
\end{equation}
where the function ${\cal S}$, is just the structure function, ${\cal F}$, 
times the compensation cut-off function.
Up until now we only considered the spectrum of primordial (ie,
generated at very early times) fluctuations 
induced by cosmic defects. In our model a Harrison-Zel'dovich spectrum is predicted just as in the simplest inflationary models. The final processed
spectrum will also be the same as for the simplest 
inflationary models.

We investigate the Gaussianity of the string-induced fluctuations as
in \cite{ASWA2}. The conclusions can easily be extended
for other defect models. In the standard cosmic string scenario
the structure function 
${\cal F}(k, \eta)$ has a turn-over scale at the network correlation length,
$k_\xi=20\, (c_{\phi}\eta)^{-1}$ \cite{ASWA1,WASA}.
At a particular time, perturbations induced on scales larger than the correlation length 
are generated by many string elements and are expected to have a
nearly Gaussian.
On the other hand, perturbations induced on smaller scales are
very non-Gaussian 
because they can be either very large within the regions where a string 
has passed by 
or else very small outside these. This allows us to roughly 
divide the power spectrum 
of cosmic-string-seeded density perturbations 
into a nearly Gaussian component generated when the string correlation length
was smaller than the scale under consideration,
and a strongly skewed non-Gaussian component generated 
when the string correlation length was larger (we call these 
the `Gaussian' and `non-Gaussian' contributions respectively). 
The ratio of this two components may be easily computed by splitting 
the structure function in (\ref{pspec}), in two parts: a Gaussian part
 ${\cal F}_{\rm g}(k,\eta) = {\cal F}(k,\eta)$ for 
$k<k_\xi$ (${\cal F}_{\rm g}=0$ for $k>k_\xi$) and a non-Gaussian part 
${\cal F}_{\rm ng}(k,\eta)= {\cal F}(k,\eta)$ for $k>k_\xi$ 
(${\cal F}_{\rm ng}=0$ for $k<k_\xi$).
We can then integrate (\ref{pspec}) with this Gaussian/non-Gaussian split, 
to compute the relative contributions to the total power spectrum. The final 
result will depend on the choice of compensation scale $k_c$. If we 
take the maximum compensation scale allowed by causality \cite{RW} 
($k_c \sim 2 \, ( c_\phi \eta)^{-1}$) 
the Gaussian contribution to the total power spectrum 
will be less than $5 \%$. In any case, the non-Gaussian contribution will 
always be smaller that the Gaussian one if, as expected, the 
compensation scale is larger or equal to the 
correlation length of the string network 
($k_c \le k_\xi$). Departures from a Gaussian 
distribution are scale independent and analogous to those of 
standard defect models on large scales.

By allowing for a characteristic velocity for the scalar field $c_\phi$ 
much larger than the velocity of light (and gravity), we were able to 
construct a model with primordial, adiabatic ($\delta_r=4 \delta_m/3$), 
nearly Gaussian fluctuations whose primordial spectrum is of the 
Harrison-Zel'dovich form. This is almost indistinguishable from 
the simplest inflationary models (as far as structure formation is concerned) 
except for the small non-Gaussian component which could be detected with 
future CMB experiments. The $C_l$ spectrum and the polarization curves 
of the CMBR predicted by this model should also be identical to the ones 
predicted in the simplest inflationary models as the perturbations in the 
CMB are not generated `directly' by the defects.

\section{Discussion and conclusions}
\label{concl}

We presented further evidence of the non-negligible overlap
between topological defect and inflationary structure formation models.
The key ingredient is having the speed of the defect-producing scalar field
much larger than the speed of gravity and standard model
particles. This provides a `violation of causality', as required by \cite{hor}.
The only distinguishing characteristic of this model, by comparison with
the simplest inflationary models, will be a small non-Gaussian signal.

Admittedly our model could be considered ``unnatural'' in the
context of our present theoretical prejudices, and the same can certainly
be said about other examples such as ``mimic inflation'' \cite{turok}
and ``designer inflation'' \cite{desi}. Be that as it may, however,
the fact that these examples can be constructed (and one wonders how many
more are possible) highlights the fact that extracting robust predictions
from cosmological observations is a much more difficult and subtle
task than many experimentalists (and theorists) believe.

\acknowledgements

We thank Paul Shellard for useful discussions and comments.
C.M. is funded by FCT (Portugal) under
`Programa PRAXIS XXI', grant no. PRAXIS XXI/BPD/11769/97.
We thank CAUP for the facilities  provided. 


\end{document}